\documentclass[journal]{IEEEtran}

\usepackage{ifpdf}

\usepackage{cite}

\ifCLASSINFOpdf
\usepackage[pdftex]{graphicx}
\else
\usepackage[dvips]{graphicx}
\fi

\usepackage[cmex10]{amsmath}

\usepackage{url}

\label{tag:packages} 

\usepackage{fancyvrb}

\usepackage[show]{todo}

\usepackage{color}

\newtheorem{corollary}{\textbf{Corollary}}

\newcommand{\qv}{\textsc{QHV}}

\begin{document}
\title{Quick Hypervolume}
\author{Lu\'{i}s~M.~S.~Russo, Alexandre~P.~Francisco
  \thanks{Both authors are with INESC-ID/KDBIO and the Department of Engenharia
    Inform\'{a}tica, Instituto Superior T\'{e}cnico - Universidade
    T\'{e}cnica de Lisboa: lsr@kdbio.inesc-id.pt,
    aplf@kdbio.inesc-id.pt }}
\markboth{Journal of \LaTeX\ Class Files,~Vol.~6, No.~1, January~2007}%
{Shell \MakeLowercase{\textit{et al.}}: Bare Demo of IEEEtran.cls for Journals}

\maketitle

\begin{abstract}
  We present a new algorithm to calculate exact hypervolumes. Given
  a set of $d$-dimensional points, it computes the
  hypervolume of the dominated space. Determining this value is an
  important subroutine of Multiobjective Evolutionary Algorithms
  (MOEAs). We analyze the ``Quick Hypervolume'' (\qv) algorithm
  theoretically and experimentally. The theoretical results are
  a significant contribution to the current state of the art. Moreover
  the experimental performance is also very competitive, compared
with existing exact hypervolume algorithms.

A full description of the algorithm is currently submitted to IEEE
Transactions on Evolutionary Computation.
\end{abstract}

\begin{IEEEkeywords}
diversity methods, hypervolume, multiobjective optimization,
performance metrics.
\end{IEEEkeywords}

\IEEEpeerreviewmaketitle
\urlstyle{tt}
\section{Introduction}
\label{sec:introduction} 
\IEEEPARstart{I}{n} this paper we focus on problems that optimize
several objectives at the same time. Most of the times these objectives
conflict with each other, meaning that maximizing one objective
implies a loss of performance in another. An illustrative example of
this problem is children's Christmas gift lists. Children are usually
not trying to maximize any particular objective, except possibly the
number of gifts, and moreover are not mindful of the overall
budget. Parents on the other hand are given the hard task of choosing
which gift, or gifts, to buy. This is no trivial task, as the number
of criteria/objectives involved is big. How much ``fun'' is the gift?
In which case games are preferable to socks. Will it help in
developing some talent? Where maybe books are preferable to games. Is
it going to be useful? How long will it be in use? What is the cost
per use? In which case, one might prefer socks. Of course children
usually do not enjoy getting socks. Since the objectives are not
measurable these problems are even harder than Multiobjective
optimization problems.

As the number of objectives and of items under analysis increases, the
complexity of the problem increases considerably. Namely the time the
problem takes to be solved, due to the large number of possible
choices. We seem to have intuitive knowledge of this complexity. From
a psychological point of view this may have the negative impact of
increasing anxiety~\cite{schwartz2005paradox}. Interestingly, as the
amount of choice increases so do the artifacts people use for coping
with complexity.

MOEAs~\cite{deb2009multi} solve multiobjective optimization problems
which occur in a wide range of problems, scheduling, economics,
finance, automatic cell planning, traveling salesman, etc. Updated
surveys on these algorithms are readily
available~\cite{zhou2011multiobjective,goh2009multi}. There is a class
of MOEAs in which we are particularly interested because they use
indicators to guide their decisions, namely they use the
hypervolume~\cite{Inc1, Inc2, Inc3, Inc4}, or the generational
distance.

We study the complexity of the algorithms that
compute hypervolumes, specifically the space and time performance. We
obtain the following results:
\begin{enumerate}
\item Section~\ref{sec:pivot-divide-conquer} describes a new, divide
  and conquer, algorithm for computing hypervolumes, \qv. The
  algorithm is fairly simple, although it requires some implementation
  details.
\item Section~\ref{sec:theoretical} includes a theoretical analysis of
  \qv. Assuming the points
are uniformly distributed on a hyper-sphere or hyper-plane, the analysis shows
  that \qv\ takes $O(d n^{1.1} \log^{d-2} n)$ time to solve an
  hypervolume problem, with $n$ points in $d$ dimensions. This bound
holds for
  the average case, and with high probability. The power in
  $n$ converges to $1$, therefore it can also be bounded by $O(d
  n^{1.01} \log^{d-2} n)$ or less.
\item We study this performance experimentally. Our \qv\ prototype is
extremely competitive against state of the art
  hypervolume algorithms, see Section~\ref{sec:testing}.
\end{enumerate}

Let us move on to the hypervolume problem.
\section{The Problem}
\label{sec:problem} 
\begin{figure}[b]
  \centering
  \immediate\write18{make axis-points.tex}
  \input{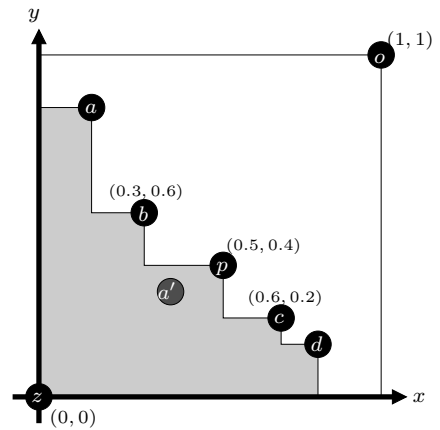}
  \caption{The area of a set of 2D points.}
  \label{fig:2DHV}
\end{figure}

Given a set of $d$-dimensional points we seek to compute
the hypervolume of the dominated space. This section gives a precise
description of the problem. Fig.~\ref{fig:2DHV} shows a set of points
and the respective 2D hypervolume, commonly referred to as area.
The region of space under consideration is delimited by a
rectangle with opposing vertexes $z$ and $o$, that are close to $0$
and $1$, respectively. We consider only rectangles that are parallel
to the axis.

We say that point $p$ dominates\footnote{A point does not
  dominate another point that has the same coordinates, this is the
  sole exception of the rectangle criterion we gave.} point $a'$
because $a'$ is contained
in the rectangle of vertexes $z$ and $p$. Notice that we cannot
state that $d$ dominates $a'$, since the rectangle with vertexes $z$
and $d$ does not contain $a'$.

Given a set of points, in our example $\{a, a', b, p, c, d\}$, we want
to compute the \textbf{dominated} area, shown in gray in
Fig.~\ref{fig:2DHV}. In general we use hyperrectangles, instead of
rectangles, to define dominance between points. The point $z$ is
always a vertex of these hyper-rectangles. The opposing vertex is a
point $p$, from the set. Therefore the problem is to compute the
hypervolume occupied by the
hyper-rectangles that use the points in the set, but without counting
the dominated space twice, just like in the 2D example.

 The coordinates can be any reals in $[0,1]$. Our algorithm uses a
sub-routine to eliminate dominated points, so we do not insist on
having a set of non-dominated points.
\section{Pivot Divide and Conquer}
\label{sec:pivot-divide-conquer} 
In this section we describe the \qv\ algorithm, by working our way
from 2D to higher dimensions and gradually introduce the necessary
concepts. Pivot divide and conquer is the technique used by
QuickSort\cite{Hoare:1961:AQ:366622.366644}. The process consists of
the following three steps:
\begin{enumerate}
\item Select a ``special'' pivot point. This point is processed and
  excluded from the recursion.
\item Divide the space according to the pivot, more precisely
  classify points into the possible space regions.
\item Recursively solve each of the sub-problems in the ``smaller''
  regions of space, and add up the hypervolumes.
\end{enumerate}
\subsection{The 2D case}
\label{sec:2d-case}
\begin{figure}[b]
  \centering
  \immediate\write18{make axis-divide.tex}
  \input{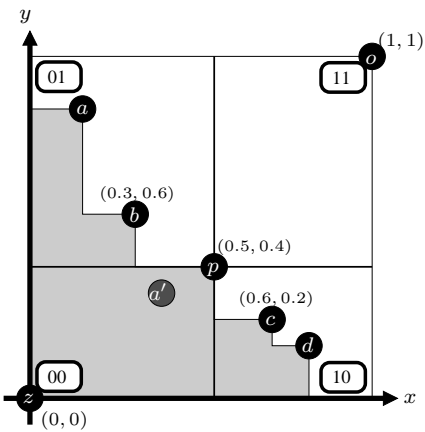}
  \caption{Pivot Divide and Conquer for 2D points. The quadrants are
    labeled by binary numbers.}
  \label{fig:PDV2D}
\end{figure}

Fig.~\ref{fig:PDV2D} shows an example of this process, in 2D. First we
choose point $p$ to be the pivot. Second we divide the rectangle, of
vertexes $z$ and $o$, according to $p$. Third we recursively compute
the area of the points in quadrants $01$ and $10$.

$\cdots$
\section{Analysis}
\label{sec:analysis} 

$\cdots$
\subsection{Theoretical}
\label{sec:theoretical}

$\cdots$
\subsubsection{Average Case}
\label{sec:average-case}

There are certainly classes of problems that can not be modeled this
way. Therefore we restrict our attention to hyper-planes and
hyper-spheres with uniform point distribution, for which the model is
appropriate.

\begin{corollary}
\label{cor:avg}
  Consider a class of $d$-dimensional hypervolume problems of points 
   uniformly distributed on a plane or on the surface of a
hyper-sphere. For any  positive number $\epsilon > 0$, the optimistic
\qv\ algorithm solves an $n$ point hypervolume
problem, from this class, in $O(d n^{1 + \epsilon}\log^{d-2}n)$ expected time.
\end{corollary}

$\cdots$
\begin{corollary}
\label{cor:high}
  Consider a class of $d$-dimensional hypervolume problems of points 
   uniformly distributed on a plane or on the surface of a
hyper-sphere. For any small positive number  $\epsilon$, such that $0
< \epsilon < 1$, the optimistic
\qv\ algorithm solves an $n$ point hypervolume
problem, from this class, in $O(d n^{1 + \epsilon}\log^{d-2}n)$ time,
with at
least $1-1/n$ probability. Formally:
\[
Pr \left(T(n) < n^{1 + \epsilon} \times O(d \log^{d-2}n) \right) \geq 1- \frac{1}{n}
\]
where $T(n)$ is the time and $c$ a constant.
\end{corollary}

$\cdots$
\subsection{Testing}
\label{sec:testing} 
We now present experimental results for estimating the overlap power
and compare our \qv\ prototype with state of the art alternatives.

\subsubsection{Experimental Setting}
\label{sec:experimental-setting}
Let us now focus on the system time and space performance of the
algorithm. In particular we will show how the techniques proposed in
the previous section affect performance. Our implementation of the \qv\
algorithm is available at \url{http://kdbio.inesc-id.pt/~lsr/QHV/}

All results where obtained on a quad-core processor at 3.20GHz, with
256KB of L1 cache, 1MB of L2 cache, 8MB of L3 cache, and 8GB of main memory. The
prototypes were compiled with \texttt{gcc} 4.7.1. For \qv\ we used
\texttt{-msse2} flag to support SSE2 and
passed the cache line size into the code
\verb|-DCLS=$$(getconf LEVEL1_DCACHE_LINESIZE)|, this was used to
align memory to cache lines. The dimension was also determined at
compile time, so there is a different binary for each dimension, this
allows for better loop unrolling. Each one of the other prototypes contains a
sophisticated makefile that produces optimized binaries, it
automatically selects the best flags for a given platform. We used the
optimized binaries produced by those makefiles, for our
platform. Besides the proper flags we included
\texttt{-s -static -m32 -O9 -march=core2}.

\begin{figure}[t]
  \centering
  \scalebox{0.24}{\includegraphics{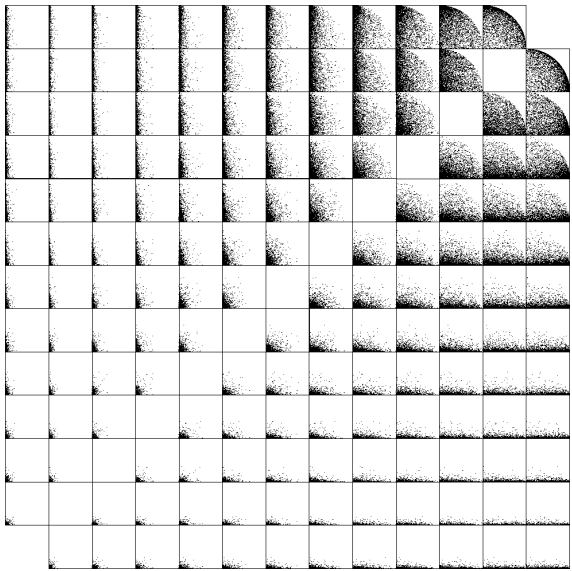}}
  \caption{A matrix of plots for the 13D spherical points of the WFG
    dataset. The file contained around 300 points, a similar
    distribution can be observed in the remaining files of the dataset.}  
  \label{fig:WFGdata}
\end{figure}

In the experimental evaluation we used the benchmarks available at
\url{http://www.wfg.csse.uwa.edu.au/hypervolume/}, by While \emph{et
  al.}~\cite{WFG}. In particular, we rely on the second benchmark with
dimensions ranging from 3 to 13, with spherical, random, discontinuous and degenerate front
datasets. Each dataset contains from 10 to 1000 points, depending on
the number of dimensions. We performed 10 runs per
dataset, each one with 20 fronts. The spherical points of the WFG dataset are not uniformly
randomly chosen on a hyper-sphere's surface with $d$ dimensions.
To
illustrate dimensions dependency, Fig.~\ref{fig:WFGdata} shows a matrix of
plots, each plot shows the points, plotted in 2D, by choosing two
coordinates and discarding the rest.

\begin{figure}[t]
  \centering
  \scalebox{0.24}{\includegraphics{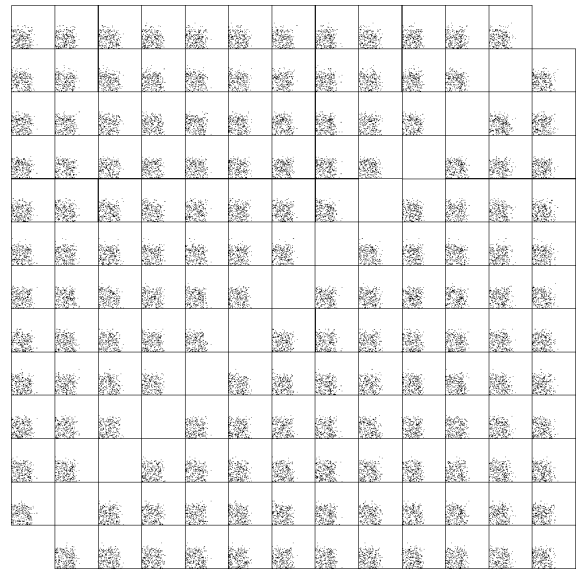}}
  \caption{A matrix of plots for the 13D spherical points of our
    dataset. The file contained 300 points.}  
  \label{fig:LSRdata}
\end{figure}

We also generated our own datasets, in such a way that the points
where chosen uniformly at random on the hyper-sphere's surface. More
precisely we generated a random point in $[0,1]^d$, using the
\texttt{drand48} function, from \texttt{stdlib.h}. These points are
uniformly random on the
given space, but not necessarily on a hyper-sphere surface. We
projected the points into the surface, by calculating the distance to
the origin and dividing every coordinate by this value. Thus obtaining
points at distance $1$. The resulting points can be seen in
Fig.~\ref{fig:LSRdata}.

\subsubsection{Experimental Performance}
\label{sec:exper-perf}
\begin{figure}[H]
  \centering
  \immediate\write18{make plotS3d.tex}
  \input{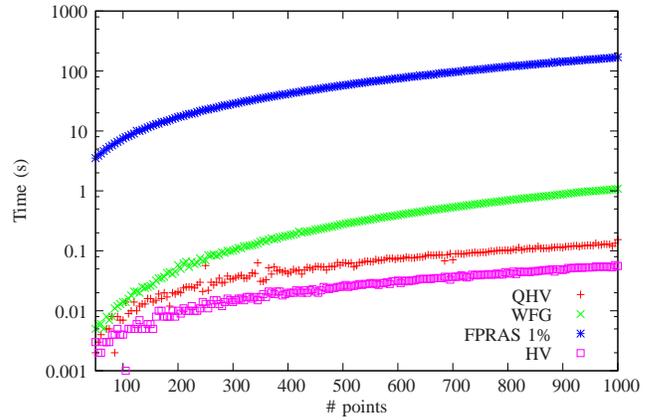}
  \caption{3D spherical fronts. }
  \label{fig:plotS3d}
\end{figure}

\begin{figure}[H]
  \centering
  \immediate\write18{make plotS4d.tex}
  \input{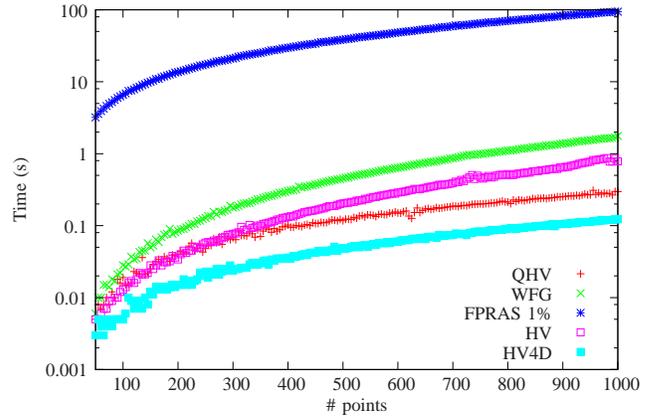}
  \caption{4D spherical fronts. }
  \label{fig:plotS4d}
\end{figure}

\begin{figure}[H]
  \centering
  \immediate\write18{make plotS5d.tex}
  \input{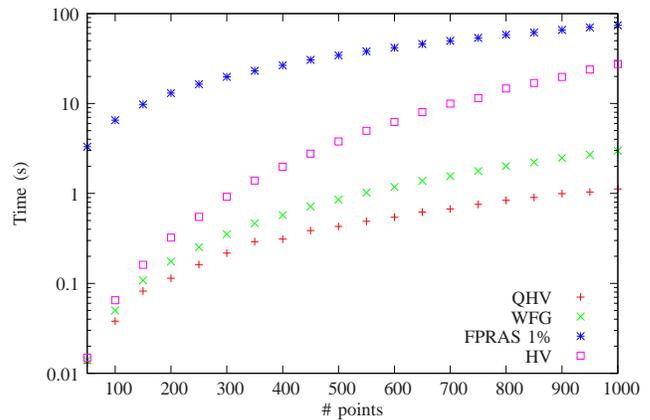}
  \caption{5D spherical fronts. }
  \label{fig:plotS5d}
\end{figure}

\begin{figure}[H]
  \centering
  \immediate\write18{make plotS6d.tex}
  \input{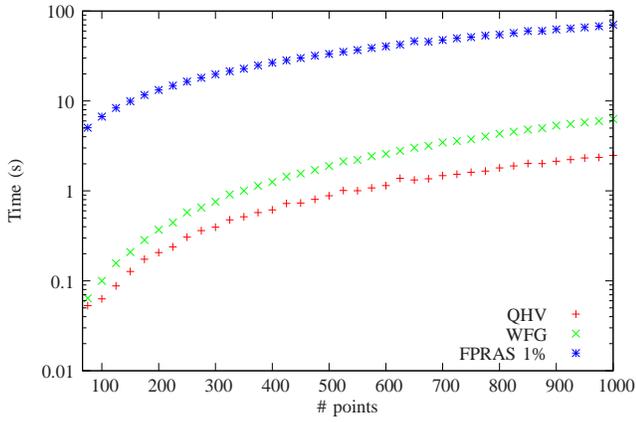}
  \caption{6D spherical fronts. }
  \label{fig:plotS6d}
\end{figure}

\begin{figure}[H]
  \centering
  \immediate\write18{make plotS7d.tex}
  \input{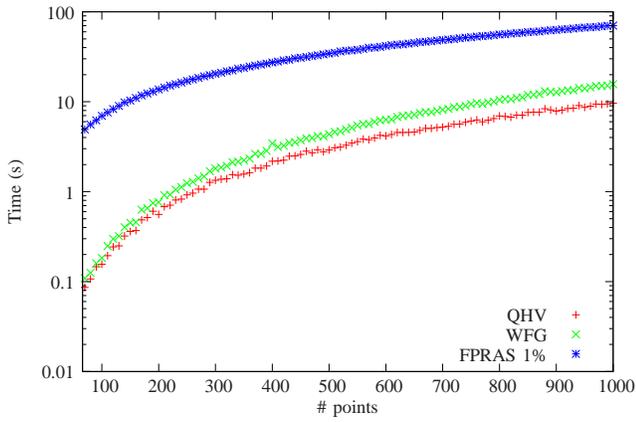}
  \caption{7D spherical fronts. }
  \label{fig:plotS7d}
\end{figure}

\begin{figure}[H]
  \centering
  \immediate\write18{make plotR4d.tex}
  \input{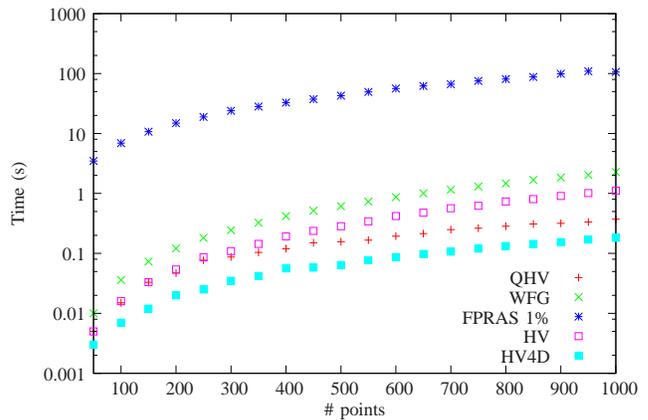}
  \caption{4D random fronts. }
  \label{fig:plotR4d}
\end{figure}

\begin{figure}[H]
  \centering
  \immediate\write18{make plotR5d.tex}
  \input{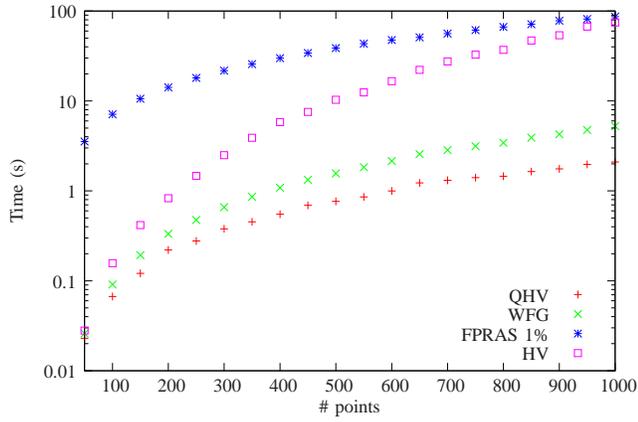}
  \caption{5D random fronts. }
  \label{fig:plotR5d}
\end{figure}

\begin{figure}[H]
  \centering
  \immediate\write18{make plotR6d.tex}
  \input{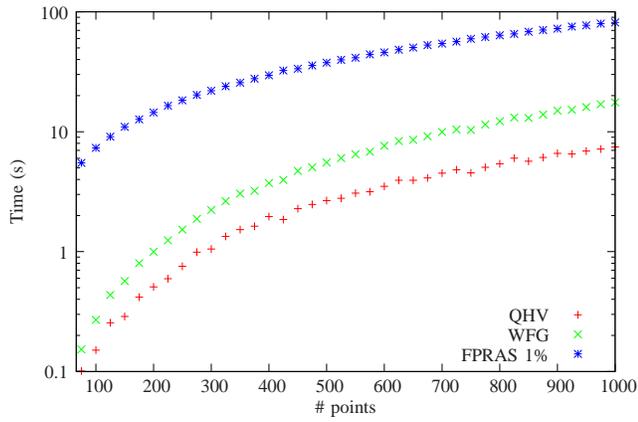}
  \caption{6D random fronts. }
  \label{fig:plotR6d}
\end{figure}

\begin{figure}[H]
  \centering
  \immediate\write18{make plotR7d.tex}
  \input{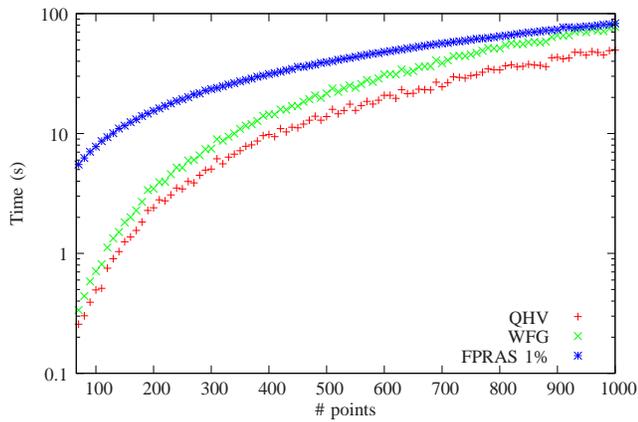}
  \caption{7D random fronts. }
  \label{fig:plotR7d}
\end{figure}

\begin{figure}[H]
  \centering
  \immediate\write18{make plotR10d.tex}
  \input{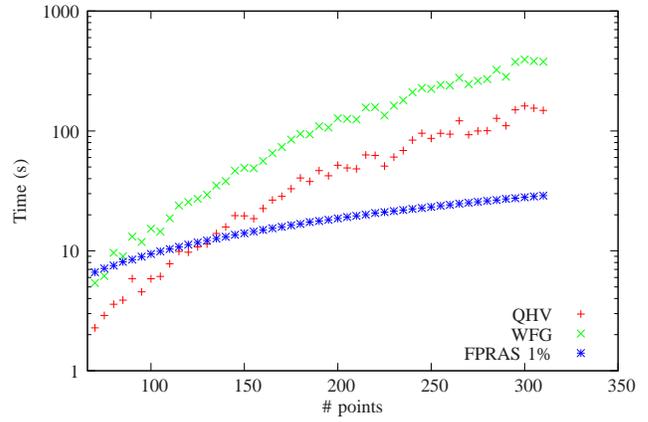}
  \caption{10D random fronts. }
  \label{fig:plotR10d}
\end{figure}

Figures~\ref{fig:plotS3d}--\ref{fig:plotC13d} show the results
concerning the running time of our \qv\ algorithm, the WFG
algorithm~\cite{WFG}, the FPRAS~\cite{bringmann2010approximating},
with $\epsilon = 0.01$, the HV algorithm, which is an improved version
of FLP (\url{http://iridia.ulb.ac.be/~manuel/hypervolume}), and an optimized
version for 4D, HV4D, provided by the
authors~\cite{DBLP:conf/cccg/GuerreiroFE12}. We use logarithmic scales to cope with
the gap in performance gap among different algorithms.

In comparing the performance of the different algorithms it is
important to point out that the FPRAS returns an $\pm \epsilon$
approximation of the hypervolume, not an exact value. Moreover the
approximation may miss this interval with 25\% probability. This
algorithm is extremely sensitive to
$\epsilon$, changing it to $0.1$ yields a $100$ times speed-up, in
practice. Therefore it is not meaningful to claim that
the FPRAS is faster or slower in a given dataset, it depends on the
precision that is required by the application. We choose a fixed
reasonable $\epsilon$ to show how the performance evolves. The real
hypervolume decreases exponentially with $d$, therefore the estimate
should become inaccurate for higher dimensions. We inspected the
resulting values and observed that estimates are actually reasonably
accurate. 
  
By far the worst performance of \qv\ occurs in the degenerated
dataset. This is not an abnormal behavior of \qv. Instead the
remaining algorithms behave much
better than usual, for that kind of data set. Like \qv\ the FPRAS
algorithm also maintains a consistent performance. Notice that for 10D the
WFG algorithm is around 1000 times faster for this dataset than it is
for the spherical dataset. Clearly \qv\ and the FPRAS are ignoring
some intrinsic property of the dataset. 

We can observe that HV is the fastest algorithm in 3D, but the
performance degrades quickly. We omitted HV
for higher dimensions, because of this slowdown. In 3D \qv\ is the
second best algorithm. In 4D the fastest
algorithm is the HV4D. \qv\ still obtains competitive
performance, usually better than HV for a large number of points,
except on the degenerated dataset. For higher dimensions, \qv\ is the
fastest algorithm and the performance becomes similar to WFG for very high 
dimensions, for example
13D, Fig.~\ref{fig:plotS13d}. This is partially a consequence of the
non-uniform dependencies on the data set. To show this fact we run a
13D test on our dataset, Figures~\label{fig:plotS3dOur} to
~\ref{fig:plotS13dOur}. The results still
show a large gap between WFG and \qv, where \qv\ remains faster. In
13D we tested less points, because the algorithms
became $100$ times slower.

 Note that at 13D the FPRAS obtains much
better performance than \qv\ and WFG, partially because we did not
applied a correction to $\epsilon$. The estimates are in fact fairly
accurate, although without a theoretical guarantee.

In Table~\ref{tab:mem} we show the memory peak requirements of the
different algorithms. It can be observed that, up to 7 dimensions, \qv\
requires less space than HV and WFG. The memory requirements of \qv\
increase with $d$, because we need to store a counter for each
hyperoctant. As we mentioned in Section~\ref{sec:analysis}, this cost
can be avoided by using a hash, which would also have an impact on the
time performance. To avoid this effect we chose not to implement the
hash. In fact the space requirements of these algorithms are modest,
for example WFG loads all the point sets in a test into memory, to
minimize this effect we reduced the tests so that they contain only
one set of points. Although the memory peak increases for \qv\ the
same happens to WFG, in 13 dimensions \qv\ also requires less space
than WFG.

 \begin{table}[b]
   \centering
\begin{tabular}[h]{*{8}{|l}|}
\multicolumn{8}{c}{Heap Peaks in KB} \\ \hline
    &  3D  &  4D  &  5D  &  6D  &  7D  &  10D  &  13D \\ \hline
QHV & 28.7 & 37.0 & 66.3 & 129.3& 242.3& 795.1 &  311.8  \\
WFG &125.1 &195.4 &273.6 & 359.6& 453.3& 781.5 &  1180.0 \\
FPRAS  & 51.7 & 59.5 & 67.3 & 75.2 & 83.0 & 107.0 & 131.1 \\
HV  &168.2 &199.5 &230.8 & 262.1&      &       &   \\ 
HV4D  &   & 121.2 &      &      &      &       & \\ 
\hline
\multicolumn{8}{c}{Stack Peaks in KB} \\ \hline
QHV &  2.4 &  2.4 &  4.8 & 4.9  &  7.1 & 12.4  &  16.3    \\ 
WFG &  9.5 &  9.5 &  9.5 & 9.5  &  9.5 &  9.5  &  9.5    \\
FPRAS  & 1.4 &  1.4 & 1.4 & 1.4 & 1.4 & 1.4 & 1.4 \\ 
HV  &  9.8 &  9.8 &  9.8 & 9.8  &      &       & \\
HV4D  &      &  9.6 &      &      &      &       & \\ 
\hline
\end{tabular}
   \caption{Memory peaks of different algorithms on spherical fronts with 1000 points.}
   \label{tab:mem}
 \end{table}

\section{Conclusions and Further Work}
\label{sec:concl-furth-work} 
In this paper we proposed a new algorithm for computing hypervolumes,
\qv. We focused on performance, time and space complexity. The \qv\
algorithm uses a divide and conquer strategy, which is different from
the usual line sweep approach. The resulting algorithm is fairly
simple and efficient. We analyzed \qv\ theoretically and
experimentally.

\subsection{Conclusions}
\label{sec:conclusions}

We expect the \qv\ algorithm to have a
significant impact in the future development of MOEAs, in that it
makes comparing more objectives feasible.

\qv\ is still devoid  of extra features. Designing a version of \qv\
that can compute exclusive hypervolumes is an important unattended
task. Other closely related problems may also benefit from the pivot
divide and conquer strategy of \qv, namely computing empirical
attainment functions~\cite{springerlink:10.1007/3-540-44719-9_15}.

\section*{Acknowledgments}
We would like to thank the Walking Fish Group for providing standard
test sets of points and answering our questions regarding the WFG
prototype, namely some subtleties on the orientation of the dominance.

We are deeply grateful to Andreia Guerreiro and Carlos Fonseca for
showing us the hypervolume problem~\cite{guerreiro2011efficient} and
providing prototypes and positive feedback. Likewise we are also
grateful to Tobias Friedrich for his enthusiasm about our initial
results, also for providing software and bibliography.

This work was supported by national funds through FCT –
Funda\c{c}\~{a}o para a Ci\^{e}ncia e a Tecnologia, under project
PEst-OE/EEI/LA0021/2011, projects TAGS PTDC/EIA-EIA/112283/2009,
NetDyn PTDC/EIA-EIA/118533/2010 and HELIX, PTDC/EEA-ELC/113999/2009.

\ifCLASSOPTIONcaptionsoff
  \newpage
\fi

\begin{IEEEbiography}[{\includegraphics[width=1in,height=1.25in
]{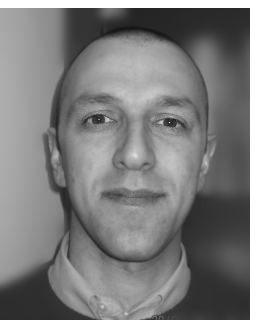}}]{Lu\'{i}s M. S. Russo}
  received the Ph.D. degree from Instituto Superior T\'{e}cnico,
  Lisbon, Portugal, in 2007. He is currently an Assistant Professor
  with Instituto Superior T\'{e}cnico. His current research
  interests include algorithms and data structures for string
  processing and optimization.
\end{IEEEbiography}
\begin{IEEEbiography}[{\includegraphics[width=1in,height=1.25in,clip,keepaspectratio]{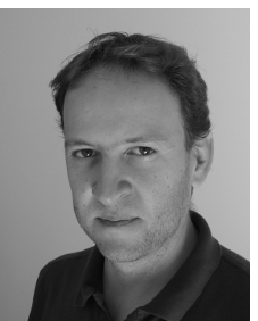}}]{Alexandre P. Francisco}
has a Ph.D. degree in Computer Science and Engineering and he is currently an Assistant Professor at the CSE Dept, IST,
Tech Univ of Lisbon. His current research interests include the design and analysis of data structures and algorithms, with applications on
network mining and large data processing.

\end{IEEEbiography}

\clearpage
\begin{figure}[H]
  \centering
  \immediate\write18{make plotS10d.tex}
  \input{plotS10d.tex}
  \caption{10D spherical fronts. }
  \label{fig:plotS10d}
\end{figure}
\begin{figure}[H]
  \centering
  \immediate\write18{make plotS13d.tex}
  \input{plotS13d.tex}
  \caption{13D spherical fronts. }
  \label{fig:plotS13d}
\end{figure}
\begin{figure}[H]
  \centering
  \immediate\write18{make plotS3dOur.tex}
  \input{plotS3dOur.tex}
  \caption{3D spherical fronts, from our dataset. }
  \label{fig:plotS3dOur}
\end{figure}
\begin{figure}[H]
  \centering
  \immediate\write18{make plotS4dOur.tex}
  \input{plotS4dOur.tex}
  \caption{4D spherical fronts, from our dataset. }
  \label{fig:plotS4dOur}
\end{figure}
\clearpage
\begin{figure}[H]
  \centering
  \immediate\write18{make plotS5dOur.tex}
  \input{plotS5dOur.tex}
  \caption{5D spherical fronts, from our dataset. }
  \label{fig:plotS5dOur}
\end{figure}
\begin{figure}[H]
  \centering
  \immediate\write18{make plotS7dOur.tex}
  \input{plotS7dOur.tex}
  \caption{7D spherical fronts, from our dataset. }
  \label{fig:plotS7dOur}
\end{figure}
\begin{figure}[H]
  \centering
  \immediate\write18{make plotS13dOur.tex}
  \input{plotS13dOur.tex}
  \caption{13D spherical fronts, from our dataset. }
  \label{fig:plotS13dOur}
\end{figure}
\begin{figure}[H]
  \centering
  \immediate\write18{make plotD3d.tex}
  \input{plotD3d.tex}
  \caption{3D degenerated fronts. }
  \label{fig:plotD3d}
\end{figure}
\begin{figure}[H]
  \centering
  \immediate\write18{make plotD4d.tex}
  \input{plotD4d.tex}
  \caption{4D degenerated fronts. }
  \label{fig:plotD4d}
\end{figure}
\begin{figure}[H]
  \centering
  \immediate\write18{make plotD5d.tex}
  \input{plotD5d.tex}
  \caption{5D degenerated fronts. }
  \label{fig:plotD5d}
\end{figure}
\clearpage
\begin{figure}[H]
  \centering
  \immediate\write18{make plotD6d.tex}
  \input{plotD6d.tex}
  \caption{6D degenerated fronts. }
  \label{fig:plotD6d}
\end{figure}
\begin{figure}[H]
  \centering
  \immediate\write18{make plotD7d.tex}
  \input{plotD7d.tex}
  \caption{7D degenerated fronts. }
  \label{fig:plotD7d}
\end{figure}
\begin{figure}[H]
  \centering
  \immediate\write18{make plotD10d.tex}
  \input{plotD10d.tex}
  \caption{10D degenerated fronts. }
  \label{fig:plotD10d}
\end{figure}
\begin{figure}[H]
  \centering
  \immediate\write18{make plotD13d.tex}
  \input{plotD13d.tex}
  \caption{13D degenerated fronts. }
  \label{fig:plotD13d}
\end{figure}

\begin{figure}[H]
  \centering
  \immediate\write18{make plotR3d.tex}
  \input{plotR3d.tex}
  \caption{3D random fronts. }
  \label{fig:plotR3d}
\end{figure}
\clearpage
\begin{figure}[H]
  \centering
  \immediate\write18{make plotR13d.tex}
  \input{plotR13d.tex}
  \caption{13D random fronts. }
  \label{fig:plotR13d}
\end{figure}

\begin{figure}[H]
  \centering
  \immediate\write18{make plotC3d.tex}
  \input{plotC3d.tex}
  \caption{3D discontinuous fronts. }
  \label{fig:plotC3d}
\end{figure}
\clearpage
\begin{figure}[H]
  \centering
  \immediate\write18{make plotC4d.tex}
  \input{plotC4d.tex}
  \caption{4D discontinuous fronts. }
  \label{fig:plotC4d}
\end{figure}
\begin{figure}[H]
  \centering
  \immediate\write18{make plotC5d.tex}
  \input{plotC5d.tex}
  \caption{5D discontinuous fronts. }
  \label{fig:plotC5d}
\end{figure}
\begin{figure}[H]
  \centering
  \immediate\write18{make plotC6d.tex}
  \input{plotC6d.tex}
  \caption{6D discontinuous fronts. }
  \label{fig:plotC6d}
\end{figure}
\begin{figure}[H]
  \centering
  \immediate\write18{make plotC7d.tex}
  \input{plotC7d.tex}
  \caption{7D discontinuous fronts. }
  \label{fig:plotC7d}
\end{figure}
\begin{figure}[H]
  \centering
  \immediate\write18{make plotC10d.tex}
  \input{plotC10d.tex}
  \caption{10D discontinuous fronts. }
  \label{fig:plotC10d}
\end{figure}
\begin{figure}[H]
  \centering
  \immediate\write18{make plotC13d.tex}
  \input{plotC13d.tex}
  \caption{13D discontinuous fronts. }
  \label{fig:plotC13d}
\end{figure}
\clearpage

\end{document}